\documentclass[12pt]{article}
\usepackage{authblk}
\usepackage{url}

\usepackage{amssymb, amsmath}
\usepackage{caption, ctable}
\usepackage[nameinlink]{cleveref}
\usepackage{colortbl}

\def\Grob   {Gr{\"{o}}bner\ }

\newcommand{\F}{\mathbb{F}}

\newcommand{\Fq}{{\mathbb{F\!}_q}}

\def\bSA         {\hbox{\bf SA}}
\def\bA         {\hbox{\bf A}}
\def\bS         {\hbox{\bf S}}

\def\bW         {\hbox{\bf W}}
\def\bC         {\hbox{\bf C}}
\def\bF         {\hbox{\bf F}}

\def\cL      {{\cal L}}

\def\cJL      {J{\cal L}}

\def\Z      {\mathbb{Z}}

\def\N      {\mathbb{N}}

\definecolor{purple}{rgb}{0.54, 0.17, 0.89}

\newtheorem{definition}{Definition}
\newtheorem{theorem}{Theorem}
\newtheorem{remark}{Remark}

\newtheorem{prop}{Proposition}
\newtheorem{example}{Example}
\newtheorem{conjecture}{Conjecture}

\newenvironment{proof}{\par\noindent\\ {\bf Proof: }}{\hfill $\Box$ \\}

\newcolumntype{?}{!{\vrule width 1.25pt}}

\newcommand{\PreserveBackslash}[1]{\let\temp=\\#1\let\\=\temp}
\newcolumntype{C}[1]{>{\PreserveBackslash\centering}p{#1}}

\newcommand{\thline}{\specialrule{.1em}{.05em}{.05em}}

\newcommand\figref[1]{\centerline{Figure~}}

\title{Analysis and Computation of Multidimensional Linear Complexity of Periodic Arrays}

\author[1]{Rafael Arce}
\author[1]{Carlos Hern\'andez}
\author[1]{Jos\'e Ortiz}
\author[1]{Ivelisse Rubio}
\author[2]{Jaziel Torres}
\affil[1]{Department of Computer Science, University of Puerto Rico, R\'{\i}o Piedras}
\affil[2]{Department of Mathematics, University of Puerto Rico, R\'{\i}o Piedras}
\date{}                     
\begin{document}

\maketitle

\abstract{Linear complexity is an important parameter for arrays that are used in applications related to information security.  In this work we  survey constructions of two and three dimensional arrays, and present new results on the multidimensional linear complexity of periodic arrays obtained using the definition and method proposed in \cite{ArCaGoMoOrRuTi,GoHoMoRu,MoHoRu}. The results include a generalization of a bound for the linear complexity, a comparison with the measure of complexity for multisequences, and computations of the complexity of arrays with periods that are not relatively prime for which the ``unfolding method'' does not work. Conjectures for exact formulas and  the asymptotic behavior of the complexity of some array constructions are formulated. We also present open source software for constructing multidimensional arrays and for computing their multidimensional linear complexity.}




\section{Introduction}

Multidimensional periodic arrays are useful in applications such as digital watermarking, multiple target recognition and communications \cite{Al,GoGoTi2019,LeMoTi,MoTi2011,MoTi2012,Ortiz2020,YaJiHe}. It is desirable to have arrays with a variety of sizes. Depending on the particular application, the array should satisfy properties such as good auto and cross correlation, balance, and complexity. Randomly generated arrays pose problems to provide properties such as periodicity and orthogonality. Precomputed arrays are stored in memory, which imposes a heavy memory burden on some systems. Hence, it is important to provide algebraic constructions for arrays that have the desired properties and are easily implemented. Several constructions have been proposed and their properties analyzed over the years.

Since some of the applications are related to information security, it is particularly important that the arrays have good complexity, meaning that they are resistant to Berlekamp-Massey types of attacks, where the complete array might be deduced from knowing some of its entries. The linear complexity of sequences has been widely studied \cite{DiHeSh,MacSl,Massey}.  However, not much work has been done on the analysis of the complexity of multidimensional arrays.  A definition of the complexity of 2-dimensional arrays viewed as multisequences was given in \cite{MeNi}. The computation of multidimensional linear complexity of 2-dimensional arrays with periods that are relatively prime was done by ``unfolding'' the array into a sequence and applying the Berlekamp-Massey algorithm in \cite{LeMoTi,MoTi2012}. A new definition and theory for the computation of multidimensional linear complexity of arrays was proposed in \cite{ArCaGoMoOrRuTi,GoHoMoRu,MoHoRu}. This definition applies to any number of dimensions, does not have the restrictions of the unfolding method, and it is more accurate than the joint linear complexity defined for multisequences.

Given that there are few sequences with known formulas for  their complexity, it is expected that formulas for the exact value of the complexity of arrays would be hard to find.  In this work we present a generalization of a bound for the linear complexity of arrays presented in \cite{ArCaGoMoOrRuTi} and conjectures for exact formulas, and  the asymptotic behavior of the complexity of some array constructions. It is also proved that the definition of multidimensional linear complexity in \cite{ArCaGoMoOrRuTi,GoHoMoRu,MoHoRu} is more accurate than the definition of joint linear  complexity of multisequences. We present new computations of the complexity of families of multidimensional arrays for wireless communications and watermarking applications presented in \cite{MoTi2011,MoTi2012} for which the complexity was unknown. In addition, we provide open source software to compute the multidimensional linear complexity of arrays of any dimension, and a web application that can be used to input or construct arrays of up to three dimensions and calculate their linear complexity.

\section{Multidimensional Linear Complexity of Periodic Arrays}

We consider periodic arrays with entries over a finite field $\F_q, \ q=p^r,\ p $ a prime, and denote the set of non-negative integers by $\N_0$. A sequence $\bS=s_0, s_1, \dots$ is a 1-dimensional array and  has period $n \in \N$ if $n$ is the smallest such that $s_{i+n}=s_i$ for all $i \in \N_0$. A polynomial $f(x)=\sum_{i \in {Supp(f)}} f_ix^i$ defines a linear recurrence relation on the sequence $\bS$ if $\sum_{i \in {Supp(f)}}f_is_{i+\beta} =0$ for all $\beta \in \N_0$, where $Supp(f)$ is the set of indices of the non-zero terms of $f$. We say that these recurrence polynomials are  {\bf valid} on the sequence $\bS$. The set of all valid polynomials on $\bS$, $Val(\bS)$, forms an ideal in $\F_q[x]$, the ring of polynomials in the variable $x$ and coefficients in $\F_q$. The {\bf linear complexity} of $\bS$, $\cL(\bS)$, is the degree of the minimal (monic) generator of $Val(\bS)$, $m(x)$, which can be found using the well-known Berlekamp-Massey algorithm. For the generalization to multiple dimensions it is important to note that $\cL(\bS)$ is also the number of monomials that are not divisible by the lead monomial of $m(x)$. Since the sequence has period $n$, the polynomial $x^n-1$ is in $Val(\bS)$ and hence $\cL(\bS) \leq n$.

A  2-dimensional infinite array over $\F_q$  is a function $\bA: \N_0^2 \rightarrow \F_q$, and we denote  $\bA(i,j)$  by $a_{ij}$. We say that $\bA$ is {\bf periodic} with period vector $(n_1, n_2)\in \N^2$ if $a_{i+n_1k_1, j+n_2k_2}=a_{i,j}$ for $k_1, k_2 \in \N_0$ and all $(i, j) \in \N_0^2$. These arrays can be represented by a subarray of dimensions $n_2 \times n_1$ and we do so by associating its entries to the integer coordinates of the first quadrant of the Cartesian plane (\Cref{coorde}).

 \begin{figure}[htbp] 
 \centering
 $\bA=\begin{array}{?c|c|c|c?}
 \thline
  a_{0,n_2-1}  & a_{1,n_2-1} &  \cdots &  a_{n_1-1,n_2-1}   \\ \hline
   &  &  \ddots &    \\ \hline
 a_{0,1}  & a_{1,1} &   &  a_{n_1-1,1}   \\ \hline
 a_{0,0}  & a_{1,0} &  \cdots &  a_{n_1-1,0}  \\ \thline
 \end{array}\ .
 $
 \caption{Labeling of the entries of an $n_2 \times n_1$ array $A$.}
 \label{coorde}
 \end{figure}


A polynomial $f(x,y)=\sum_{i,j \in {Supp(f)}} f_{i,j}x^iy^j$ defines a {\bf linear recurrence relation on the array} $\bA$ if $\sum_{i,j \in {Supp(f)}}f_{i,j}a_{i+\beta_1, j+ \beta_2} =0$ for all $\beta_1, \beta_2 \in \N_0$. We say that these polynomials  are {\bf valid on the array $\bA$}.  The set of all valid polynomials on $\bA$, $Val(\bA)$, forms an ideal in $\F_q[x,y]$, the ring of polynomials in the variables $x,y$ and coefficients in $\F_q$. This ideal might not be generated by a single polynomial but it has  finite generating sets. In particular, $Val(\bA)$ is generated by a \Grob basis with respect to a monomial ordering $ \leq_T$ that can be computed using Sakata's algorithm or the Rubio-Sweedler-Taylor algorithm described in \cite{RuSwHe}. We restricted our description to 2-dimensional arrays in order to simplify the notation but the previous discussion applies to higher dimensions. 

Let $\Delta_{Val(\bA),\leq_T} $ denote the  set of exponents of all monomials that do not occur as leading monomials in $Val(\bA)$ with respect to $ \leq_T$. As a result of the \Grob bases properties, $\Delta_{Val(\bA), \leq_T}$ is also the set  of exponents of all monomials that are not divisible by any lead monomial in a \Grob basis for $Val(\bA)$ with respect to $ \leq_T$, and hence can be computed from the \Grob basis.  The size of $\Delta_{Val(\bA),\leq_T}$, denoted $\lvert\Delta_{Val(\bA),\leq_T}\rvert,$ is the dimension of $\Fq[x_1, \ldots, x_m]/Val(\bA)$ as a $\Fq$-vector space and hence it is invariant under monomial orderings. We just write $\lvert\Delta_{Val(\bA)}\rvert$ for the size of this set.

\begin{definition} \label{defLinComp} Let $\bA$ be a multidimensional
  periodic array and $Val(\bA)$ be the ideal of  linear recurrence relations
  valid on the array. Define the {\bf multidimensional linear
    complexity} $\cL(\bA)$ of the array $\bA$ as the size of the delta 
  set of $Val(\bA)$ with respect to any monomial order; this is, $\ \cL(\bA)= \lvert \Delta_{Val(\bA)}\rvert$.
\end{definition} 

 The multidimensional linear complexity $\cL(\bA)$ can be obtained by computing a \Grob basis for $Val(\bA)$ using the Rubio-Sweedler-Taylor algorithm \cite{RuSwHe} and determining $\lvert\Delta_{Val(\bA)}\rvert$  \cite{ArCaGoMoOrRuTi}.

 If the $m$-dimensional array $\bA$ has period $(n_1,\ldots, n_m)$, the polynomials $x_1^{n_1}-1, \ldots, x_m^{n_m}-1$ are in $Val(\bA)$ and hence $\lvert\Delta_{Val(\bA)}\rvert \leq n_1n_2\cdots n_m$. With this we can define the {\bf normalized linear complexity} of the array as $\cL_n(\bA)=\cL(\bA)/\left(n_1n_2\cdots n_m\right)$, a measure that  allows us to compare the complexity of arrays of different dimensions and periods.

\subsection{Other measures for complexity}

A well studied definition for the complexity of a $2$-dimensional array with period $\left(n_1,n_2\right)$ is given by considering the array as an {\bf $n_1$-fold multisequence}, a sequence $\bS$ of sequences $\bS_1, \ldots, \bS_{n_1}$ with period $n_2$. The {\bf joint linear complexity} of $\bS,\ \cJL(\bS)$, is the degree of a minimal polynomial that is valid for each $\bS_i$. 

The linear complexity of some of the $2$-dimensional arrays presented in \cite{LeMoTi,MoTi2012} was computed by ``unfolding'' the array using the Chinese Remainder Theorem in order to construct a sequence, and then compute the complexity of the resulting sequence using the Berlekamp-Massey algorithm. This method has the limitation that the periods of the array must be relatively prime. This constrain is why the complexity of some of the arrays in those papers could not be computed (see \Cref{CompNuevas}). 

 As we will see in \Cref{ExampleMulti} on \Cref{Sec:TheoResults}, \Cref{defLinComp} is more accurate than the joint linear complexity definition because the later might miss relations among the entries of different columns;  \Cref{defLinComp} can also be used in higher dimensions. Our method is consistent with the unfolding method \cite{ArCaGoMoOrRuTi,GoGoTi2017} but the dimensions of the array do not need to be relatively prime. Our approach allows for the computation of the complexity of any multidimensional periodic array, advancing the complexity analysis of multidimensional arrays.

\section{Constructions of multidimensional arrays}

In \Cref{CompNuevas} we present families of arrays from \cite{LeMoTi,MoTi2011,MoTi2012} for which the linear complexity could not be computed using the unfolding method. To make this paper self contained, we  start by presenting a survey of the  constructions of these arrays. For the sake of simplicity we only consider 2 and 3 dimensional arrays but the methods can be extended to higher dimensions. Some of the constructions use  the $p \times p$ index table $\bW$ of a finite field $\F_{p^2}$, with respect to a primitive element $\alpha$ of $\F_{p^2}$. The entries of $\bW$ are defined by $w_{i,j} = k$ if $\alpha^k = i\alpha+j$. Since 0 is not a power of $\alpha$, an $*$ is placed as the entry $w_{0,0}$ (Figure \ref{Legendres}).

\subsection{Two dimensional Legendre arrays}

\begin{figure}[htbp] 
 \centering
 \includegraphics[width=80mm]{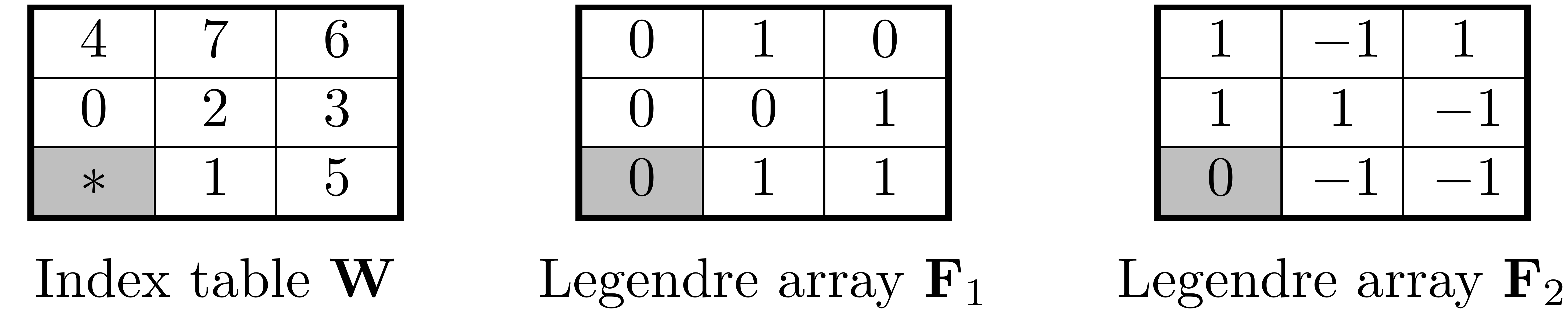}
 
 \medskip
 
 \caption{Index table and Legendre arrays corresponding to $\F_{3^2}$.}
 \label{Legendres}
 \end{figure}

A {\bf binary $2$-dimensional Legendre array} $\bF_1$ with period $(p,p)$ is constructed from an index table $\bW$ for the finite field $\F_{p^2}$ by setting $f_{0,0}=0$ and taking all other entries of $\bW$ modulo 2 (\cite{BoAnSc,MoTi2011}). Similarly, a {\bf ternary $2$-dimensional Legendre array} $\bF_2$ with period $(p,p)$ is constructed from  $\bW$  by setting $f_{0,0}=0$ and mapping the even entries of $\bW$ to 1 and the odd entries to $-1$ (\cite{MoTi2011}). For example, the arrays in \Cref{Legendres} show the index table $\bW$, the binary ($\bF_1$) and the ternary ($\bF_2$) $2$-dimensional Legendre arrays corresponding to $\F_{3^2}$, with $\alpha$ a primitive root of $x^2+2x+2$.

This construction produces solitary Legendre arrays but they will be used by the composition method as ``floors'' to construct families of 3-dimensional arrays (see \Cref{SecFloor}).

\subsection{The composition method}

\begin{figure}[htbp]
\centering
\includegraphics[width=60mm]{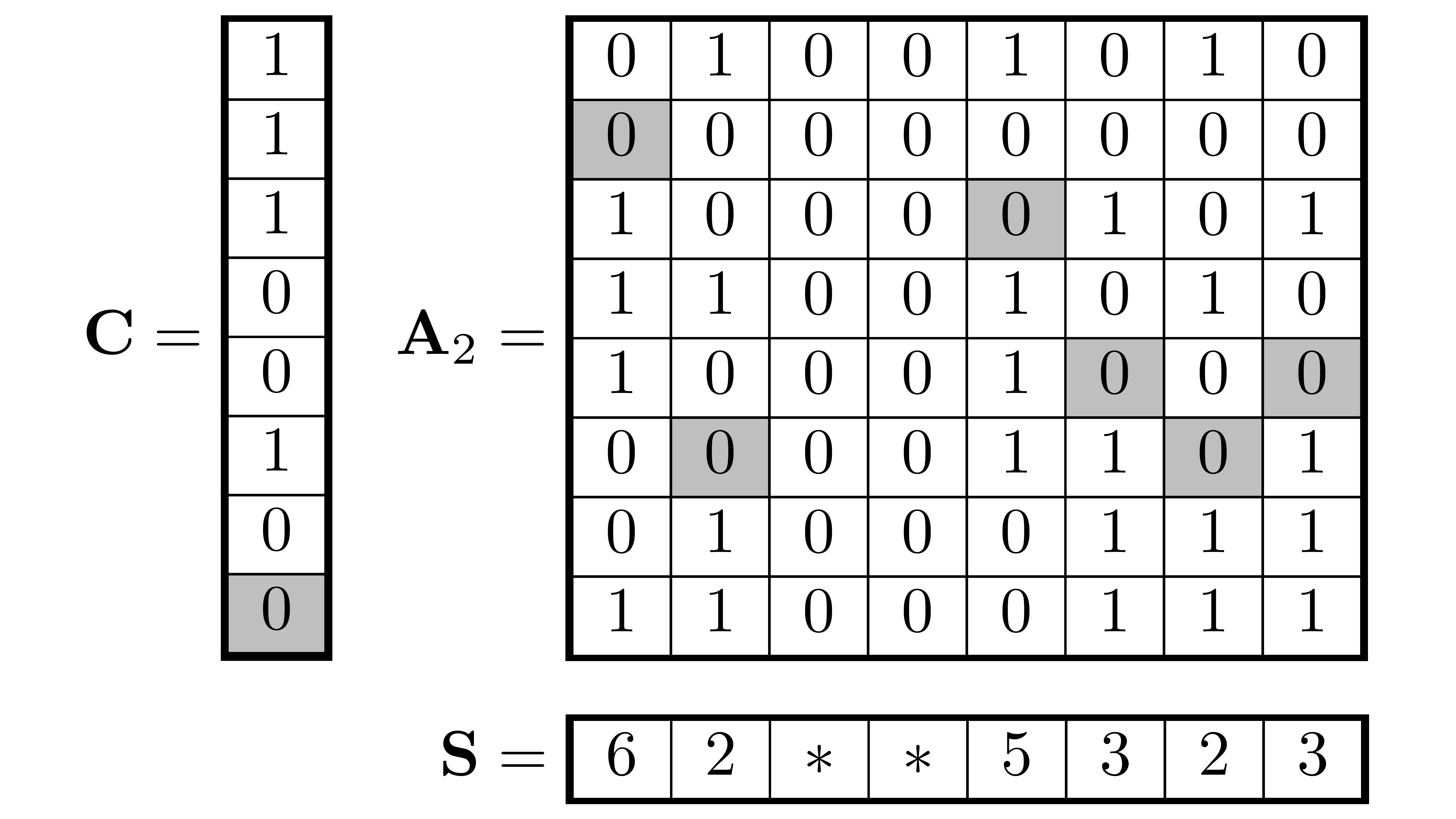}
\caption{Array $\bA_2$ is constructed by composing the logarithmic quadratic shift sequence $\bS$ with the Sidelnikov column sequence $\bC$ and $0,\ldots, 0$ in the columns corresponding to $*$.}
\label{QuadSidel}
\end{figure}

Multidimensional arrays can be constructed by composing a shift sequence/array with a column sequence or an  array of suitable  dimension \cite{MoTi2011,TiOsHa}. For example, a {\bf 2-dimensional array} $\bA$ with period  vector $(n_1,n_2)$ can be constructed using a shift sequence $\bS$ with entries in $\Z_{n_2}$ and period $n_1$ to define circular shifts of columns defined by a sequence $\bC$ of period $n_2$.  The value $s_i$ of the shift sequence $\bS$ determines the vertical cyclic shift of the column sequence $\bC$ that will be placed in the positions $(i,\_)$ of the array $\bA$: $a_{i,j}=c_{j-s_i \pmod{n_2}}$.  One can think of the value $s_i$ as the place where the first entry of the column $\bC$ will be placed (see \Cref{QuadSidel}). The entries of the shift sequence might also contain an extra symbol  for an ``undetermined shift'', i.e. the entries can be in $\Z_{n_2} \cup \{*\}$. In this case, the columns corresponding to an undetermined shift $*$ will consist of a sequence of a constant value.

\begin{figure}[htbp]
\centering
\includegraphics[width=60mm]{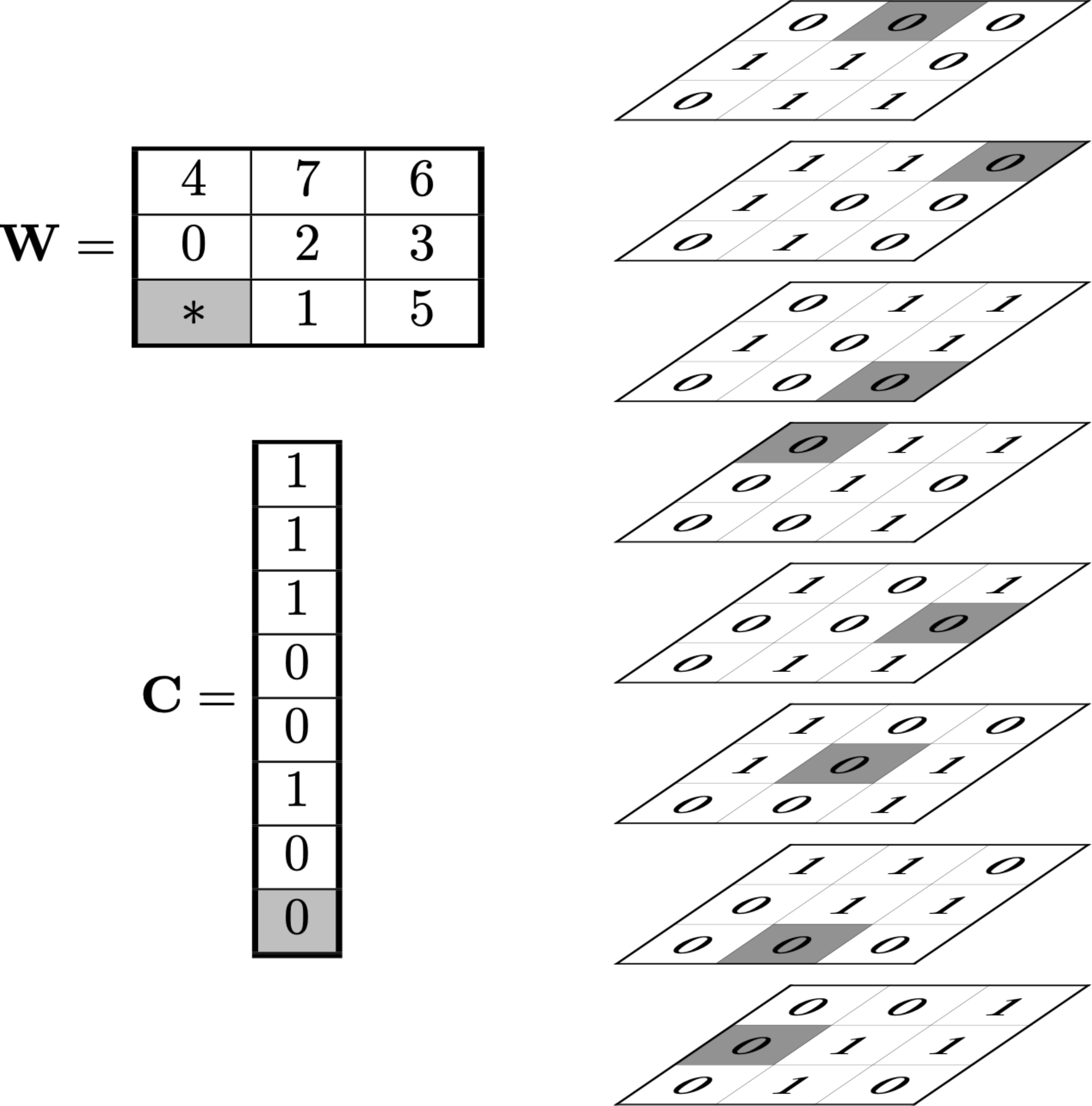}
\medskip
 \caption{ Array $\bA_3$ constructed composing the shift array from the index table, $\bSA = \bW$,  with  the Sidelnikov column $\bC$.}
\label{ShiftArraySidel}
\end{figure}

A {\bf 3-dimensional array} $\bA$ with period vector $(n_1,n_2,n_3)$ can be constructed using a 2-dimensional shift array  $\bSA$  with entries in $\Z_{n_3} \cup \{*\}$  and period vector $(n_1,n_2)$ to define circular shifts of columns given by a  column sequence $\bC$ of period $n_3$ (\cite{MoTi2011}).  The value $s_{i,j}$ of the shift array $\bS$ determines the vertical cyclic shift of the column sequence $\bC$ that will be placed in the positions $(i,j, \_)$ of the array $\bA$: $a_{i,j,k}=c_{k-s_{i,j} \pmod{n_3}}$.  One can think of the value $s_{i,j}$ as the ``floor'' of $\bA$ where the first entry of the column $\bC$ will be placed (see Figure \ref{ShiftArraySidel}). Again, the columns corresponding to an undetermined shift $*$ will consist of a column sequence of a constant value.

\begin{figure}[htbp]
\centering
\includegraphics[width=60mm]{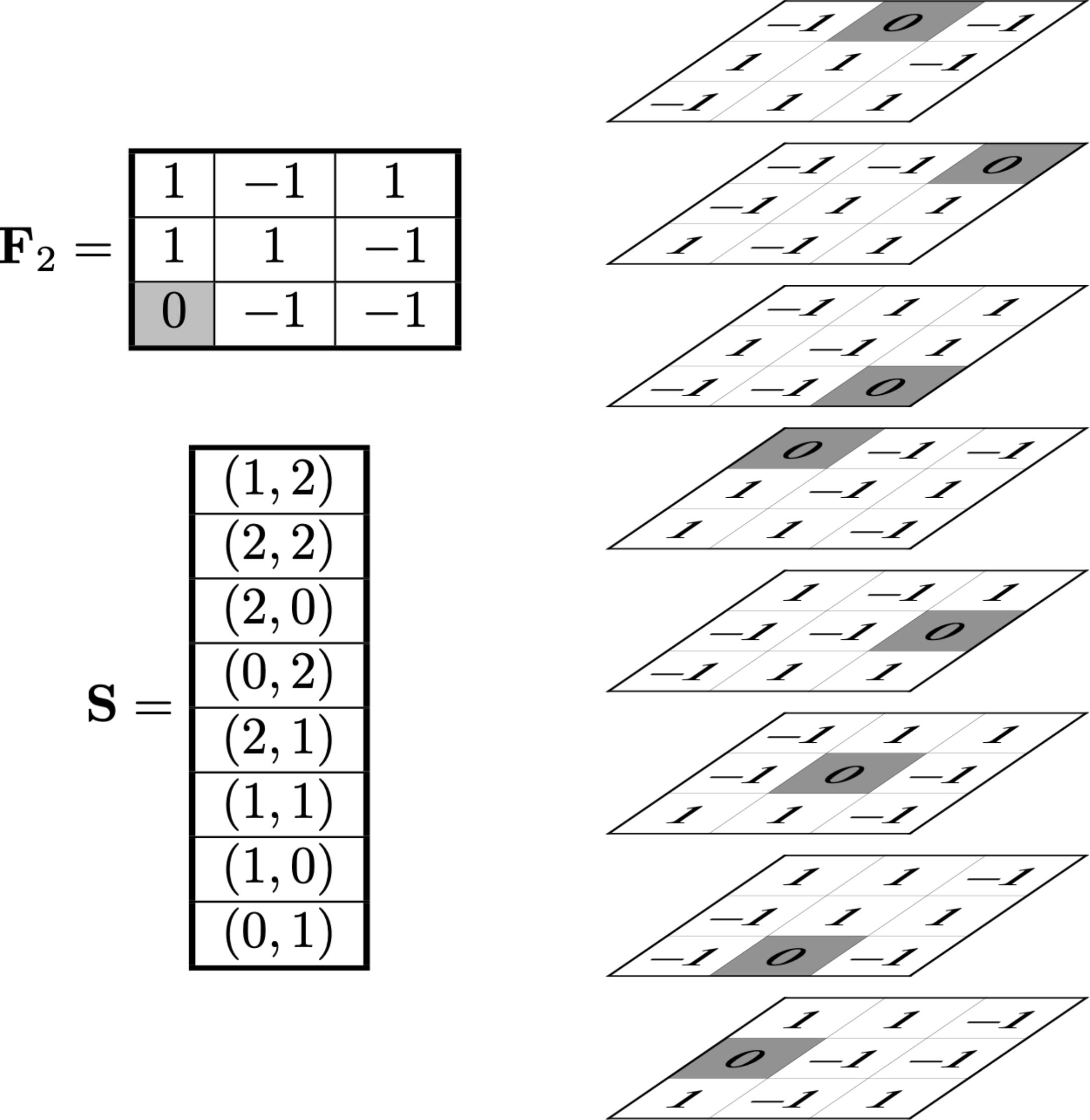}
\medskip
\caption{Array $\bA_5$ constructed composing  vector shift sequence $\bS$ with ternary Legendre floor  array $\bF_2$.  The vector shift sequence $\bS$ is obtained from the array $\bW$ in Figure \ref{Legendres}.}
\label{CoordTerLeg}
\end{figure}

Similarly,  a {\bf 3-dimensional array} $\bA$ with period vector  $(n_1,n_2,n_3)$ can be constructed by using a shift sequence $\bS$  with entries in $\Z_{n_1} \times \Z_{n_2}$   and period  $n_3$ to define circular shifts in both dimensions of ``floors'' defined by 
an array $\bF$ with period $(n_1,n_2)$ (\cite{MoTi2011}).  The value $s_k=(i,j)$ of the shift sequence $\bS$ determines the vertical and horizontal cyclic shifts of the floor array $\bF$ that will be placed in the $k$-th ``floor'' of the array $\bA$: $a_{i,j,k}=f_{(i,j)-s_k}$, where $(i,j)-s_k$ is taken modulo $(n_1,n_2)$.  One can think of the value $s_k=(i,j)$ as the ``place'' of $\bA$ where the ``first'' entry $f_{0,0}$ of the floor array $\bF$ will be placed in ``floor'' number $k$ (see Figure \ref{CoordTerLeg}).

 Note that a shift sequence $\bS$ with entries in $\Z_{n_1} \times \Z_{n_2}$   and period  $n_3$ can be obtained from a shift array $\bSA$ of dimensions $n_1 \times n_2$ and entries in $\Z_{n_3}$ all different, by assigning $s_k=(i,j)$ if $sa_{i,j}=k$.


\subsubsection{Shift sequences/arrays} 

Some of the shift sequences that can be used to construct 2-dimensional arrays are:  exponential quadratic, logarithmic quadratic, and Moreno-Maric sequences. 

To define an {\bf exponential quadratic  shift sequence over $\Z_p$} with period $p-1$, consider a quadratic polynomial $f(x)=ax^2+bx+c \in \Z_p[x],\ a \neq 0$, a primitive element $\alpha$ in $\Z_p$, and take the values of $f$ in $p-1$ consecutive powers of $\alpha$: $\bS=f\left(\alpha^0\right), f\left(\alpha^1\right), \ldots, f\left(\alpha^{p-2}\right)$ (\cite{MoTi2012}). For example, for $\alpha= 3 \in \Z_7$, the quadratic polynomial $f(x)=x^2+x+1 \in \Z_{7}[x]$ gives the exponential quadratic sequence $\bS=f\left(\alpha^0\right),  \ldots, f\left(\alpha^{5}\right)= 3, 6, 0, 1, 0, 3 $.


One can also use quadratic polynomials $f(x)$ over $\F_q$ to define shift sequences of period $q-1$ but, since we want the sequence to have entries in $\Z_{n}$, we map the values of $f(x)$ to $\Z_n$  by considering the elements of ${\F_q}^*$ as powers of a primitive element $\alpha$ of $\F_q$ and
taking their logarithm. The {\bf logarithmic quadratic shift sequence over $\F_q$} with period $q-1$ is defined by writing the non-zero values in $f\left(\alpha^0\right), f\left(\alpha^1\right), \ldots, f\left(\alpha^{q-2}\right)$ as  $f\left(\alpha^i\right)=\alpha^j$, and letting $s_i=\log_{\alpha}\left(f\left(\alpha^i\right)\right)=\log_{\alpha}\left(\alpha^j\right)$  $ = $  $j$. If $f\left(\alpha^i\right)=0$, set $s_i=*$ (\cite{MoTi2011}). For example, the quadratic polynomial $f(x)=x^2+x+2\alpha$ over $\F_{3^2}$, where $\alpha^2=\alpha+1$, has values $f\left(\alpha^0\right),  \ldots, f\left(\alpha^{7}\right)$ $=$  $\alpha^3, \alpha^6, \alpha^2, 0, 0, \alpha^5, $ $\alpha^3,$ $\alpha^2$ and gives the logarithmic quadratic shift sequence $\bS= 6, 2, *, *, 5, 3, 2, 3$ used in array $\bA_2$ of \Cref{QuadSidel}.

Under certain conditions on $\alpha$, the composition $f^n(x)$ of a rational function $f(x)=\alpha/(x+1)$ over $\F_p$ with itself, evaluated in $\F_p \cup \{\infty\}$, produces a cycle of length $p+1: 0, f^1(0), f^2(0), \cdots, f^{p-1}(0), f^p(0)=\infty$, \cite{GoGoTi2017,MoMa,MoTi2012,TiHa}\footnote{Although the conditions cited in these papers are not correct (for example, consider $x^2+x+6 \in \F_{11}[x]$ and note that the sequence obtained does not have length 12), it is true that sequences of length $q+1$ can be constructed using rational functions over $\F_q$.}.  The {\bf Moreno-Maric  shift sequence over $\Z_p$} with period $p+1$ is $\bS=0,  f^1(0)=\alpha, \ldots,  f^{p-1}(0)=-1, *$. For example, $\bS=0, 3, 2, 1, -1,*$ is the Moreno-Maric  shift sequence over $\Z_5$ with $f(x)=3/(x+1)$. Since the $*$ is always at the end of $\bS$, one can remove it to obtain a {\bf shortened Moreno-Maric  shift sequence} of length $p$. 

A {\bf Moreno-Maric  shift sequence over $\F_q$} can be constructed by writing the nonzero values of the cycle  as powers of the primitive element, $0, f^1(0)=\alpha^{i_1}, f^2(0)=\alpha^{i_2}, \cdots, f^{q-1}(0)=\alpha^{i_{q-1}}, f^{q}(0)=\infty$ and letting $\bS=*, {i_1}, {i_2}, \cdots, i_{q-1}, *$. 

To construct 3-dimensional arrays $\bA$ with period vector $(p,p, p^2-1)$ one can use an {\bf index table} $\bW$ for the finite field $\F_{p^2}$, which has entries in $\Z_{p^2-1}$ and $w_{0,0}=*$,  as shift array. The column placed in position $(0,0)$, that is, all entries $a_{0,0,k}$, will be a sequence of a constant value  (see Figure \ref{ShiftArraySidel}). 

Other 3-dimensional arrays with period vector $(p,p, p^2-1)$ can be constructed  using a {\bf vector shift sequence} $\bS$, where $s_k=(i,j)$ and $\alpha^k=i\alpha+j$, obtained from an index table $\bW$ for the finite field $\F_{p^2}$. For example, the index table $\bW$ in \Cref{Legendres} produces the  vector shift sequence $\bS=(0,1),(1,0),(1,1),(2,1),(0,2),(2,0),(2,2),(1,2)$  used in the construction of array $\bA_5$ in Figure \ref{CoordTerLeg}.

\subsubsection{Column sequences and ``floor'' arrays} \label{SecFloor} 

A good option for column sequences of period $p>2$ are {\bf Legendre  column sequences} with respect to $\Z_p$, which are defined as  $c_i=0$ if $i=0$ or $i$ is a nonsquare mod $p$,  and $c_i=1$ otherwise. For example, $\bC= 0,1,1,0,1,0,0$ is the Legendre sequence with respect to $\Z_7$. {\bf Sidelnikov  column sequences} with respect to $\F_q, \ q$ odd, are defined  as  $c_i=1$ if $\alpha^i+1$  is a nonsquare  in $\F_q$, where $\alpha$ is a primitive element, and $c_i=0$ otherwise. These sequences can be used as columns of length $q-1$, where $q$ is odd (\cite{MoTi2011}). For example, $\bC= 0,0,1,0,0,1,1,1 $ is the Sidelnikov column sequence with respect to $\F_{3^2}$ and $\alpha^2=\alpha+1$  used in array $\bA_2$ of \Cref{QuadSidel}.

The {\bf $2$-dimensional Legendre arrays} obtained from the  index table  $\bW$ for a finite field $\F_{p^2}$ (such as the ones in \Cref{Legendres}) can be used as floor arrays with period vector $(p,p)$ by letting $w_{0,0}=0$ (\cite{MoTi2011}).  With this type of floor arrays one can construct 3-dimensional arrays as it is done in the example in Figure \ref{CoordTerLeg}.

\subsubsection{Constructions for which the complexity was unknown} \label{CompNuevas}

The unfolding method was used in \cite{MoTi2011,MoTi2012} to compute the multidimensional linear complexity of several constructions. However,  this method
cannot be used to compute the complexity of some arrays described in the same papers. For example, it cannot be used for arrays of dimensions $p \times p$ such as $\bF_1$: 2-dimensional binary Legendre arrays,  $\bF_2$: 2-dimensional ternary Legendre arrays  (Figure \ref{Legendres}), or $\bA_1$:  arrays obtained from  shortened Moreno-Maric shift sequences composed with Legendre column sequences. The unfolding method can neither be used to compute the linear complexity of arrays of dimensions $(q-1) \times (q-1)$  such as  $\bA_2$:  arrays obtained from logarithmic quadratic shift sequences composed with Sidelnikov column sequences  (\Cref{QuadSidel}).

The linear complexity  of 3-dimensional arrays with dimensions $p \times p \times (p^2-1)$ such as  $\bA_3$:  arrays obtained by composing index table shift arrays with a Sidelnikov column sequences (Figure \ref{ShiftArraySidel}),   $\bA_4$: arrays obtained from vector shift sequences  composed with 2-dimensional binary Legendre floor arrays or $\bA_5$:  arrays constructed by composing vector shift sequences with  2-dimensional ternary Legendre shift arrays (\Cref{CoordTerLeg}) cannot be computed using the unfolding method.

\section{Results on complexity}

\subsection{Theoretical results} \label{Sec:TheoResults}

The following result generalizes a bound for the complexity of arrays presented as Theorem 1 in \cite{ArCaGoMoOrRuTi} to include shift sequences with unknown values. This is, shift sequences with elements in $\Z_{n_2} \cup \{*\}$. Define array $\bA$ by
$$a_{i,j}=\left\{ \begin{array}{ll}
c_{j-s_i \pmod{n_2}}, & s_i \neq * \\
0, & s_i=*
\end{array}
\right.$$

\begin{theorem} \label{Cota} Let $\bS$ be a shift sequence over $\Z_{n_2} \cup \{*\}$ with period $n_1$, $\bC$ be a column sequence over $\F_q$ with period $n_2$, and $\bA$ be the 2-dimensional array constructed with the composition method where the column corresponding to $*$ consists of $0$'s. Then, $\cL_n(\bA) \leq \cL_n(\bC)$, where $\cL_n(\cdot)$ is the normalized linear complexity.
\end{theorem}

\begin{proof} Let $m(y)$ be the minimal polynomial of $\bC$, $m'(x,y)=\sum_{j \in Supp(m)}m'_{0,j}y^j$ $=m(y)$, and $\gamma=\left(\gamma_1, \gamma_2\right)\in \N_0^2$. If $s_{\gamma_1}\neq *$, then, since $m \in Val(\bC)$ implies that $\sum_{j \in Supp(m)}m_jc_{j+\beta}=0$ for all $\beta \in \N_0$, we have
$$\sum_{(0,j) \in Supp(m')}m'_{0,j}a_{(0,j)+\gamma}= \sum_{j \in Supp(m)}m_ja_{\gamma_1,j+\gamma_2}$$ $$
=\sum_{j \in Supp(m)}m_jc_{j+\gamma_2-s_{\gamma_1}} = \sum_{j \in Supp(m)}m_jc_{j+\beta} =0,$$
where $\beta=\gamma_2-s_{\gamma_1}$, and the indices of $\bC$ are considered modulo $n_2$.

If $s_{\gamma_1}= *$, then $a_{\gamma_1, j+\gamma_2}=0$, and 
$$\sum_{(0,j) \in Supp(m')}m'_{0,j}a_{(0,j)+\gamma}= \sum_{j \in Supp(m)}m_ja_{\gamma_1,j+\gamma_2}=0.$$ Hence, for any $\gamma\in \N_0^2, \sum_{(0,j) \in Supp(m')}m'_{0,j}a_{(0,j)+\gamma}=0$ and $m'(x,y)=m(y)\in Val(\bA)$. This implies that $\Delta_{Val(\bA)}$ cannot contain exponents  of monomials that are multiples of $y^{\deg(m)}$. Since $\bS$ has period $n_1, x^{n_1}-1 \in Val(\bA)$ and $\Delta_{Val(\bA)}$ cannot contain exponents  of monomials that are multiples of $x^{n_1}$. Therefore, 
$\cL_n(\bA)=\cL(\bA)/(n_1n_2) \leq n_1\deg(m)/(n_1n_2)=\cL_n(\bC)$.
\end{proof}

\begin{remark}The above proposition is also true for 3-dimensional arrays.\end{remark}

 The selection of the values in the  column corresponding to the undefined shift ($*$) affects the complexity of the array as we can see in the next example.

\begin{example} \label{EjMoreno-Maric}
Consider the $5 \times 6$ array $\bA$ constructed by composing the Moreno-Maric  shift sequence over $\Z_5$, $\bS = 0, 3, 2, 1, 4, *$ with the Legendre  column sequence $\bC = 0,0,1,1,0$.  If the column corresponding to the undetermined shift $*$ is replaced with a column of constant $1$'s, then $\cL_n(\bA) = 13/15$, while $\cL_n(\bC) = 4/5$.
In this case $\cL_n(\bA) \nleq \cL_n(\bC)$.
\end{example}

The following refinement for the bound in Theorem 1 for the cases where $y-1$ divides the minimal polynomial of the sequence $\bC$ and $\bS$ is a shift sequence over $\Z_{n_2}$, was proved in \cite{ArCaGoMoOrRuTi}.

\begin{prop} \label{MejorCota} Let $\bS$ be a shift sequence over $\Z_{n_2}$ with period $n_1$, $\bC$ be a column sequence over $\F_q$ with period $n_2$ and minimal polynomial $m(y)$, and $\bA$ be the 2-dimensional array constructed with the composition method. If $y-1$ divides $m(y)$,  then
$\cL_n(\bA) \leq \cL_n(\bC)- \frac{n_1-1}{n_1n_2}$, where $\cL_n(\cdot)$ is the normalized linear complexity.
\end{prop}

\subsubsection{Comparison of the linear complexity of an array $\bA$ with the joint linear complexity of $\bA$ as a multisequence}

As it was mentioned before, our definition of multidimensional linear complexity is more accurate than the joint linear complexity for multisequences. 

\begin{prop} Let $\bA$ be a periodic 2-dimensional array with period $\left(n_1,n_2\right)$. Then, the normalized linear complexity of $\bA$, $\cL_n(\bA)$, is smaller or equal than the normalized joint linear complexity of $\bA$ considered as a multisequence, $\cJL_n(\bA)$. This is,
$\cL_n(\bA) \leq \cJL_n(\bA)$.
\end{prop}

\begin{proof} Let $m(y)$ be the joint minimal polynomial of an $n_1$-fold multisequence $\bA$. Then, $m(y)$ is valid for each of the columns $a_{k,0}, a_{k,1}, \ldots, a_{k,n_2-1}, \ 0 \leq k < n_1$, and $\sum_{j \in Supp(m)} m_ja_{k,j+\gamma_2}=0$ for each $0 \leq k < n_1$ and all $\gamma_2 \in \N_0$, where $j+\gamma_2$ is considered modulo $n_2$. Let $\gamma=\left(\gamma_1, \gamma_2\right) \in \N_0^2$ and $m'(x,y)=\sum_{j \in Supp(m)}m'_{i,j}x^iy^j$, where $m'_{i,j}=m_j$ for a fixed $0 \leq i < n_1$. Then, 

$$\sum_{(i,j) \in Supp(m')} m'_{i,j}a_{(i,j)+\gamma}= \sum_{i=0}^{n_1-1} \sum_{j \in Supp(m)} m_ja_{i+\gamma_1,j+\gamma_2}= \sum_{i=0}^{n_1-1} 0=0,$$ since $k=i+\gamma_1$ is fixed in the inner sum. This implies  $m'(x,y)\in Val(\bA)$, and $\Delta_{Val(\bA)}$ cannot contain  exponents of monomials that are multiples of $x^iy^{\deg(m)}$ for any $0 \leq i < n_1$.  Since $\bA$ has period $n_1$ in its first coordinate, $x^{n_1} -1 \in Val(\bA)$ and $\Delta_{Val(\bA)}$ cannot contain exponents of monomials that are multiples of $x^{n_1}$. Hence, $\cL(\bA)=\lvert \Delta_{Val(\bA)}\rvert \leq n_1\deg(m)$ and $\cL_n(\bA) \leq \deg(m)/n_2=\cJL_n(\bA)$.
\end{proof}

There are examples of arrays $\bA$ for which $\cL_n(\bA)$, is strictly smaller than $\cJL_n(\bA)$. When an array $\bA$ constructed with columns  that are shifts of the same column sequence $\bC$ is considered as a multisequence, the  minimal polynomial of $\bC$ is valid for all the columns. Hence, the normalized joint linear complexity  of $\bA$ is equal to the normalized linear complexity of $\bC,\ \cJL_n(\bA)=\cJL_n(\bC)$.  From \Cref{MejorCota}, when $y-1$ divides the minimal polynomial of $\bC$ one has $\cL_n(\bA) < \cJL_n(\bA)$. The joint linear complexity of $\bA$ misses some relations among entries of different columns.

\begin{example}\label{ExampleMulti} Consider the $7 \times 6$ array $\bA$ constructed by composing the exponential quadratic  shift sequence $\bS=3, 6, 0, 1, 0, 3$ with the  Legendre  column sequence  with respect to $\Z_7$, $\bC= 0,1,1,0,1,0,0$. Definition \ref{defLinComp} gives normalized linear complexity  $\cL_n(\bA)=19/42$. If  $\bA$ is considered as a multisequence, the normalized joint linear complexity is $\cJL_n(\bA)=4/7$ which is larger than $\cL_n(\bA)$. \end{example}

\subsection{Computational results and conjectures} \label{Computations}

Our computational results focus  on arrays for which the linear complexity could not be computed  with the unfolding method in \cite{LeMoTi,MoTi2012} because the periods of the arrays were not relatively prime. We computed the complexity of 2-dimensional $p \times p$ arrays from constructions $\bF_1,\bF_2$ and 3-dimensional $p \times p \times (p^2-1)$ arrays from constructions $\bA_3,\bA_4, \bA_5$. The complexity was computed using a C++ implementation \cite{TorresSoftware}  of the RST algorithm \cite{RuSwHe}. 
All the examples satisfy the conjectured formulas for the normalized linear complexity in \Cref{TablaFormula}.

\begin{table}[htbp]
\caption{Conjectured formulas for the normalized linear complexity.}
 \label{TablaFormula}
 \medskip
 
 \centering
{

\small \renewcommand{\arraystretch}{1.2}
\begin{tabular}{|c|c|c|} \hline
    Construction & Conjectured ${\cal L}_n(\cdot)$ & Verified for \\ \hline
    ${\bf F}_1$ & ${\frac{1}{2} - \frac{1}{2p^2}}$ & $p \leq 251$\\ [.5mm] \hline
    ${\bf F}_2$ & ${1 - \frac{1}{p^2}}$  & $3 < p \leq 109$ \\ [.5mm] \hline
    ${\bf A}_3$ & ${\cal L}_n({\bf C})[1 - \frac{1}{p^2}]$ & $p \leq 19$ \\ [.5mm] \hline
    ${\bf A}_4$ & $\frac{1}{2} - \frac{1}{2p^2}={\cal L}_n({\bf F}_1)$ & $p \leq 23$ \\ [.5mm] \hline
    ${\bf A}_5$ & $1 - \frac{1}{p^2}={\cal L}_n({\bf F}_2)$  & $3 <p \leq 17$ \\ [.5mm] \hline
\end{tabular}

}
\end{table}


Recall that  construction $\bA_3$ uses  Sidelnikov column sequences $\bC$, $\bA_4$ uses  $\bF_1$ as floor array, and $\bA_5$ uses  $\bF_2$ as floor array.
As seen in \Cref{TablaFormula}, both $\bA_4$ and $\bA_5$ have the same normalized linear complexity as their corresponding floor array.
In the case of $\bA_3$, one can see that, as the size of the array increases (size depends on $p$), the value $\cL_n(\bA_3)$ approaches the value of $\cL_n(\bC)$.

\begin{figure}[htbp]
    \centering
    \includegraphics[width=\textwidth]{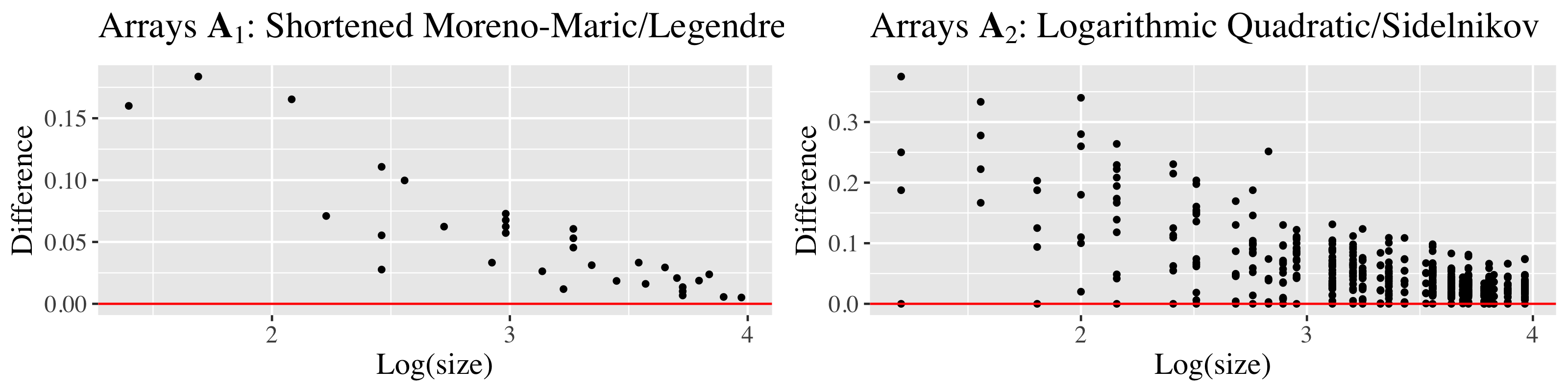}
    \caption{Difference of normalized linear complexities  $\cL_n(\bC)-\cL_n(\bA)$ as a function of the log of the size of $\bA$, $\log(n_1n_2)$, where each dot represents an array $\bA$ with period vector $(n_1,n_2)$.}
    \label{GrafA4}
\end{figure}

We do not have a conjecture  of a formula for the complexity of arrays from constructions $\bA_1, \bA_2$. However, these arrays have the same behaviour of arrays from constructions $\bA_3, \bA_4, \bA_5$, in the sense that their complexities approach the complexity of the  column/floor sequence/array. This can be seen in \Cref{GrafA4},
where the graphs show that the difference of the normalized linear complexity of  shift arrays composed with column sequences and the normalized linear complexity of the column approaches 0 as the size of the array increases.

Based on the above results and the results from \cite{LeMoTi,MoTi2012}, we have the following general conjecture.

\begin{conjecture} \label{conjetura}
If $\bA$ is an array constructed by composing a shift sequence/array with a column sequence $\bC$ or floor array $\bF$ of suitable dimensions, then as the size of $\bA$ increases, $\cL_n(\bA)$ approaches the normalized linear complexity of the column $\cL_n(\bC)$ or of the floor $\cL_n(\bF)$.
\end{conjecture}

\begin{figure}[htbp]
    \centering
    \includegraphics[width=80mm]{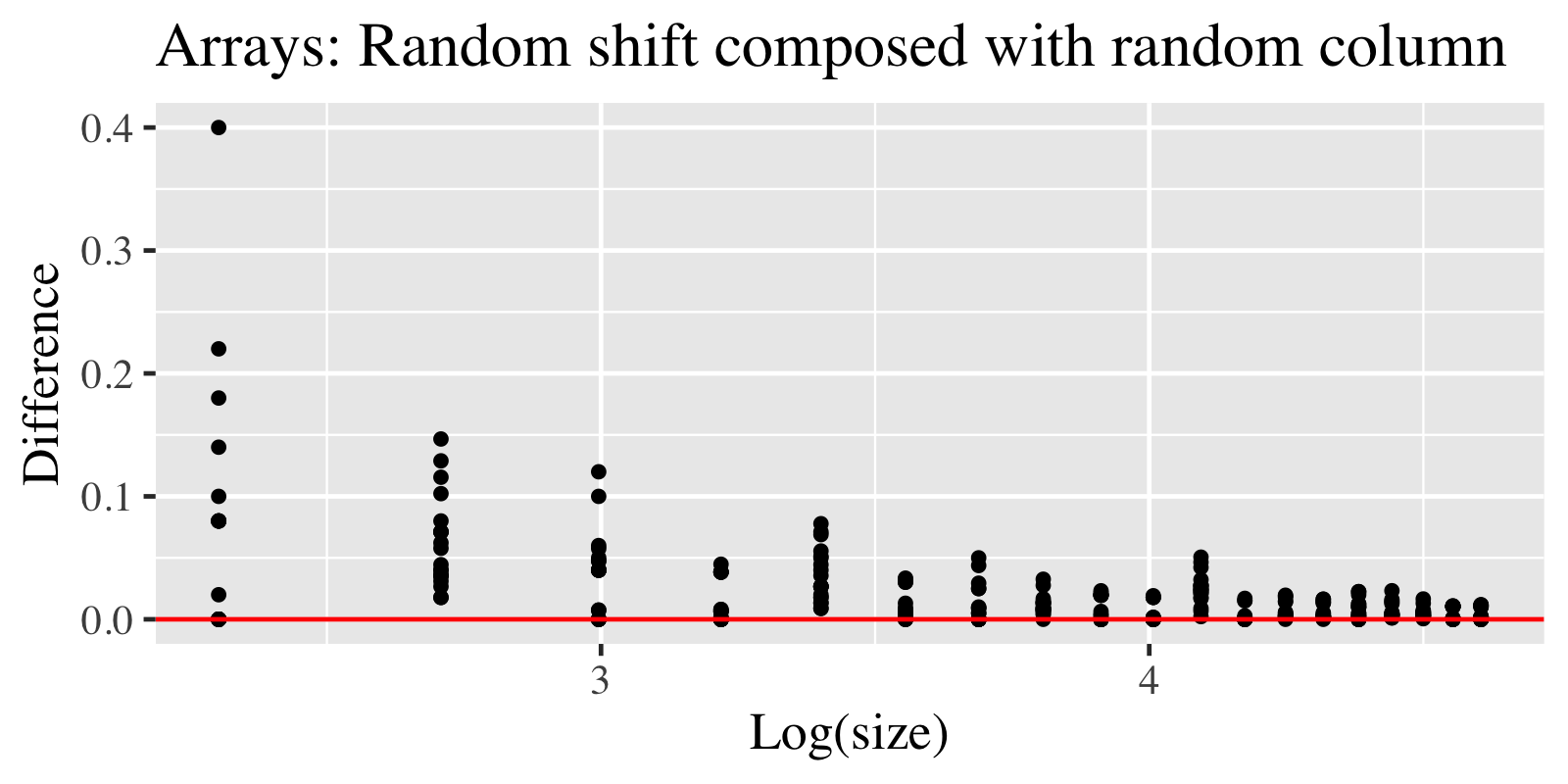}
   \caption{Difference of normalized linear complexities  $\cL_n(\bC)-\cL_n(\bA)$ as a function of the log of the size of $\bA$, $\log(n^2)$, where each dot represents an array $\bA$ with period vector $(n,n)$. Here the shift and column sequences are randomly generated.}
    \label{GrafRandom}
\end{figure}

We validated Conjecture 1 by computing the normalized linear complexity of arrays constructed by composing a randomly generated shift sequence with a randomly generated binary column sequence, each of period $n$,  for $n$ a multiple of $5, \ 5\leq n \leq 100$. We sampled 25 arrays for each $n$ and computed the normalized linear complexity of each corresponding random column sequence. In \Cref{GrafRandom} we plotted the difference between the normalized linear complexity of the random column $\bC$ and the normalized linear complexity of the array $\bA$.
We can see that, once again, as the size of the arrays increase, the difference of the normalized complexities approaches zero, validating \Cref{conjetura}.

\section{Web-based Linear Complexity Calculator}

 We have made available, as open source, the software that we developed to obtain the results in this paper (\url{https://github.com/jazieltorres/RST_Complexity}). The program is written in C++ and uses the Rubio-Sweedler-Taylor algorithm for computing a \Grob basis for ideal of linear recurrence relations on a periodic array \cite{RuSwHe} and obtain the multidimensional linear complexity of the array \cite{ArCaGoMoOrRuTi}. It supports the construction of arrays with the methods mentioned in this paper and the computation of the linear complexity of arrays up to 11 dimensions. 

 In order to facilitate the use of our program, we designed a public web  interface to the back-end software that performs such computations (\url{https://labemmy.ccom.uprrp.edu}). Hence, it allows for many of the features that the back-end software offers, such as the construction of arrays using the methods mentioned in this paper. Furthermore, it provides additional functionalities such as the generation and computation of random sequences or arrays, as well as allowing for user defined sequences and arrays.

\begin{figure}[htbp]
\centering
\begin{tabular}{cc}
{\includegraphics[width=7cm]{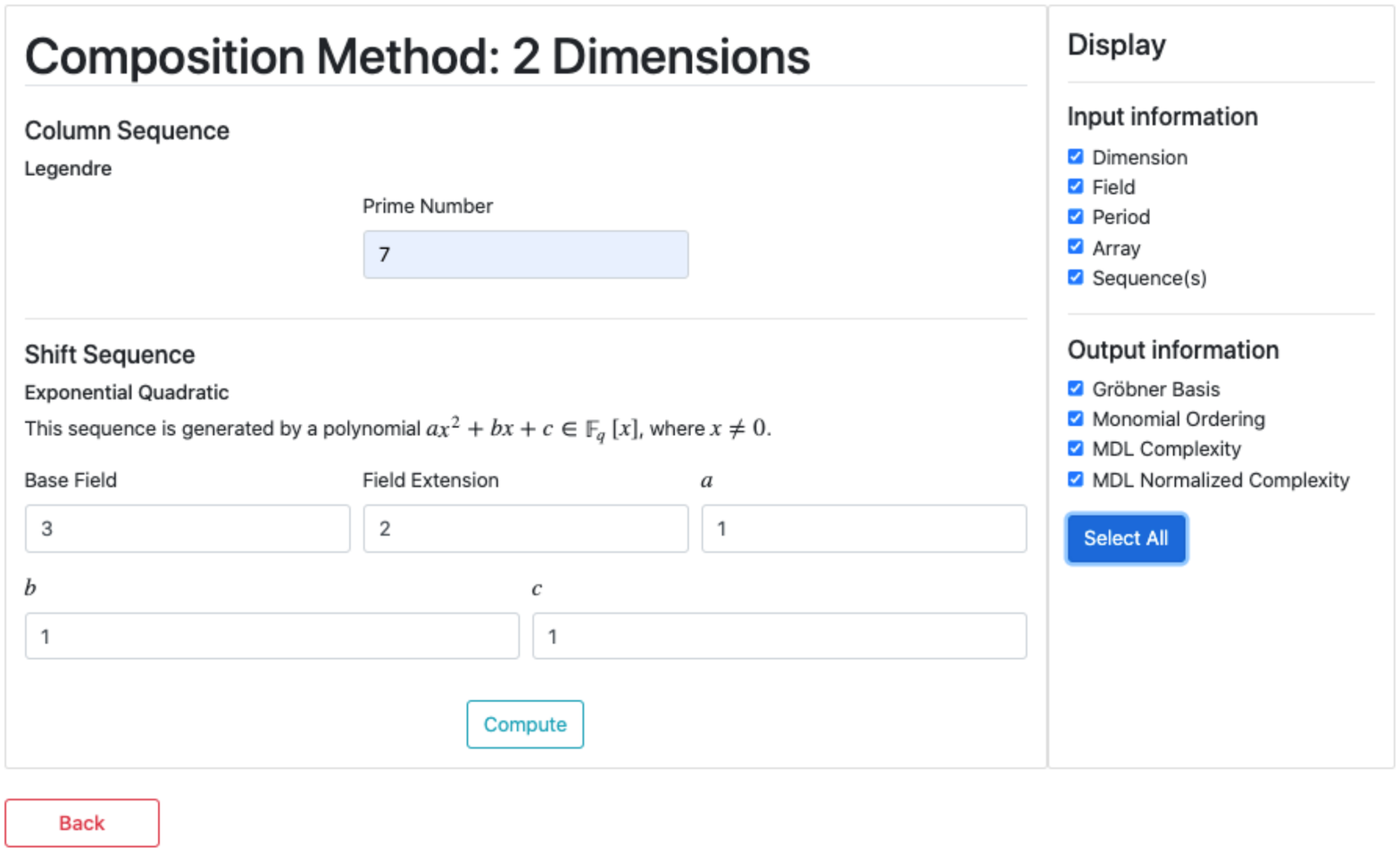}} & {\includegraphics[width=7cm]{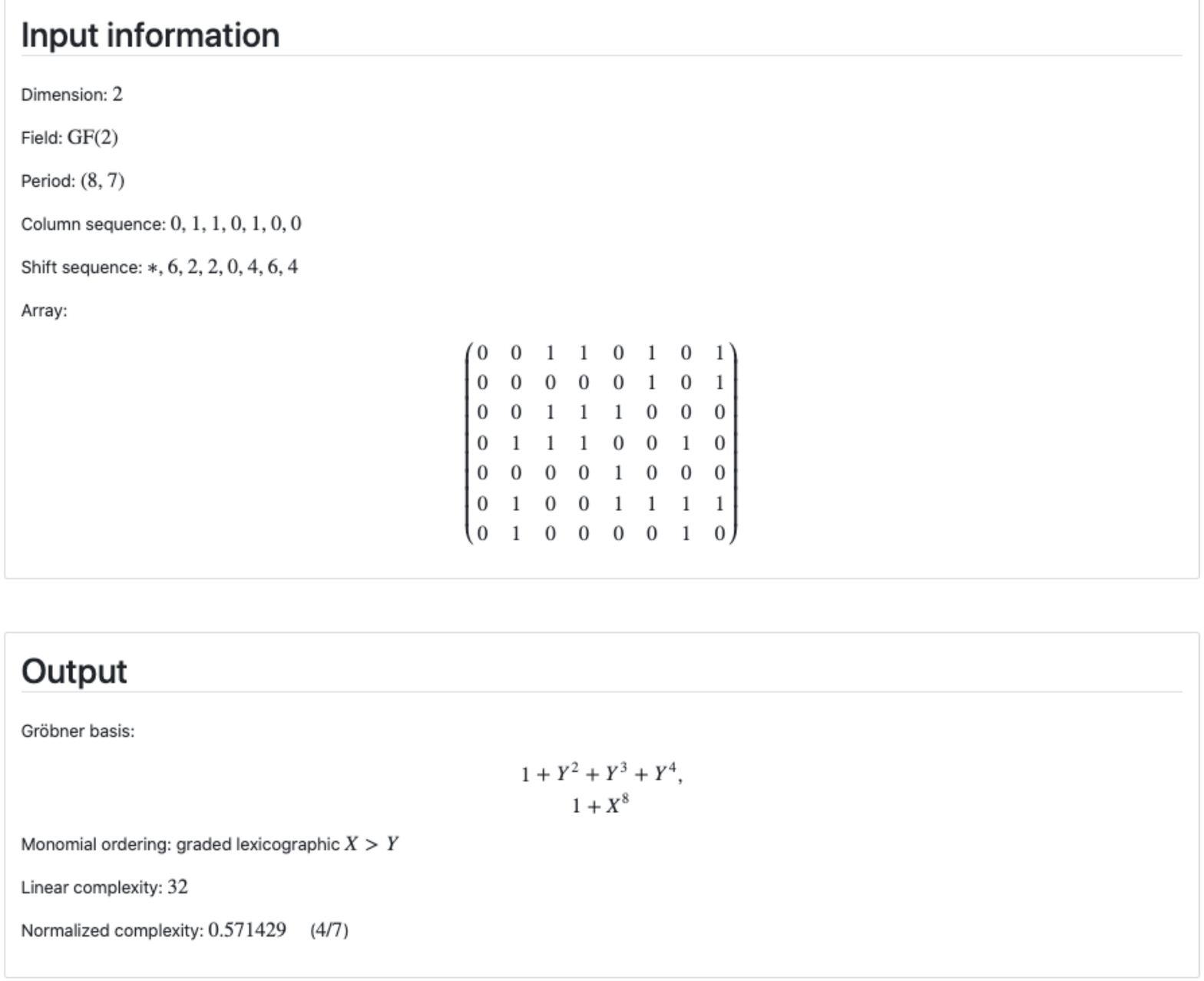}} 
\end{tabular}
\caption{Examples of input and output pages of the web application for the construction of arrays and computation of their linear complexity.}
\label{Interface}
\end{figure}

Currently, the web application supports the linear complexity computation of arrays for dimensions 1, 2 and 3. Different methods for specifying the input array are available for each dimension. Once an input method is selected and its required information submitted, a new page is rendered containing both the input information supplied and the output, the latter displaying results such as the linear and normalized linear complexities and a \Grob basis for the ideal of valid polynomials on the array. 

The online application supports computations for 1 dimensional arrays (sequences) of up to 1000 entries, 2 dimensional arrays up to 625 entries and 3 dimensional arrays up 550 entries. For larger arrays the user can use the open source software mentioned above, or download from the application a virtual machine with all the needed software.

\section{Conclusions} We reviewed definitions and methods related to the multidimensional linear complexity of periodic arrays. Constructions of arrays,  especially some for which the complexity was unknown, were surveyed. Definitions of linear complexities were compared and our definition was proved to be more accurate or general than previous definitions. A
generalization for a bound for the linear complexity was proved.
Computational results on the complexity of several constructions for which the complexity was unknown were summarized and conjectures for formulas for their exact value were presented. Our  calculations also led us to conjecture that the normalized linear complexity of arrays constructed by composing a shift sequence/array with a suitable column sequence or floor array approaches the normalized linear complexity of the column sequence or floor array.

\section*{Acknowledgment} This research was funded by the ``Fondo Institucional Para la Investigaci\'on (FIPI)'' from the University of Puerto Rico, R\'{\i}o Piedras.

\bibliographystyle{plain}

\bibliography{ConstrucComplexityMDArrays}

\end{document}